\documentclass[a4paper,11pt]{article}
\pdfoutput=1 

\usepackage{jinstpub} 
\usepackage{graphicx}
\usepackage{siunitx}
\usepackage{hyperref}
\usepackage{enumitem}
\usepackage{makecell}
\usepackage{pdfpages}

\title{\boldmath A complete simulation of the X-ARAPUCA device for detection of scintillation photons}

\author[a,1]{L. Paulucci,\note{Corresponding author.}}
\author[b]{F. Marinho,}
\author[c]{A.A. Machado}
\author[c]{and E. Segreto}

\affiliation[a]{Universidade Federal do ABC,\\Av. dos Estados, 5001, Santo Andr\'e, SP, 09210-170, Brazil}
\affiliation[b]{Universidade Federal de S\~ao Carlos,\\
Rodovia Anhanguera, km 174, 13604-900, Araras, SP, Brazil}
\affiliation[c]{Instituto de F\'isica Gleb Wataghin, Universidade Estadual de Campinas - Unicamp,\\
Rua Sergio Buarque de Holanda, No 777, CEP 13083-859 Campinas, SP, Brazil}

\emailAdd{laura.paulucci@ufabc.edu.br}

\abstract{The concept of the ARAPUCA device is relatively new and involves increasing the effective area for photon collection of SiPMs by the use of a box with highly reflective internal walls, wavelength shifters, and a dichroic filter to allow the light to enter the box and not the leave it. There were a number of tests showing the good performance of this device. Recently an improvement on the original design was proposed: the inclusion of a WLS bar inside the box to guide photons more efficiently to the SiPMs. We present a full simulation of the device using Geant4. We have included all the material properties that are available in the literature and the relevant detailed properties for adequate photon propagation available in the framework. Main results include estimates of detection efficiency as a function of the number, shape, and placing of SiPMs, width of the WLS bar, its possible attenuation, and the existence of a gap between the bar and the SiPMs. Improvement on the efficiency with respect to the original ARAPUCA design is 15-40\%. The ARAPUCA simulation has been validated in a number of experimental setups and is a useful tool to help making design choices for future experiments devices.}

\keywords{Detector modelling and simulations I (interaction of radiation with matter, interaction of photons with matter, interaction of hadrons with matter, etc); Noble liquid detectors (scintillation, ionization, double-phase); Photon detectors for UV, visible and IR photons (solid-state)  (PIN diodes, APDs, Si-PMTs, G-APDs, CCDs, EBCCDs, EMCCDs, CMOS imagers, etc).}

\proceeding{LIGHT DETECTION IN NOBLE ELEMENTS (LIDINE 2019)\\
  August 28th - 30th, 2019\\
  University of Manchester, Manchester, UK}

\begin{document}
\maketitle
\flushbottom

\section{Introduction}
\label{sec:intro}

The ARAPUCA concept \cite{arapuca} was proposed as a photon trapping device in order to increase the effective area for photon collection of silicon photomultipliers (SiPMs) for detecting LAr scintillation light. It is based on assembling a box with highly reflective internal walls and an optical window composed of a dichroic filter and two wavelength shifters (WLS), allowing the light to be converted by the first layer of shifter and pass the dichroic filter. When the photon gets converted by the second shifter, it will not be allowed to pass the filter on the way out of the box, being efficiently trapped inside it. The ARAPUCA has been tested in different experimental setups \cite{lnls, tallbo} showing a detection efficiency on the order of a few percent. 

\begin{figure}[htbp]
\begin{center}
    \centering
    \includegraphics[width=0.49\textwidth]{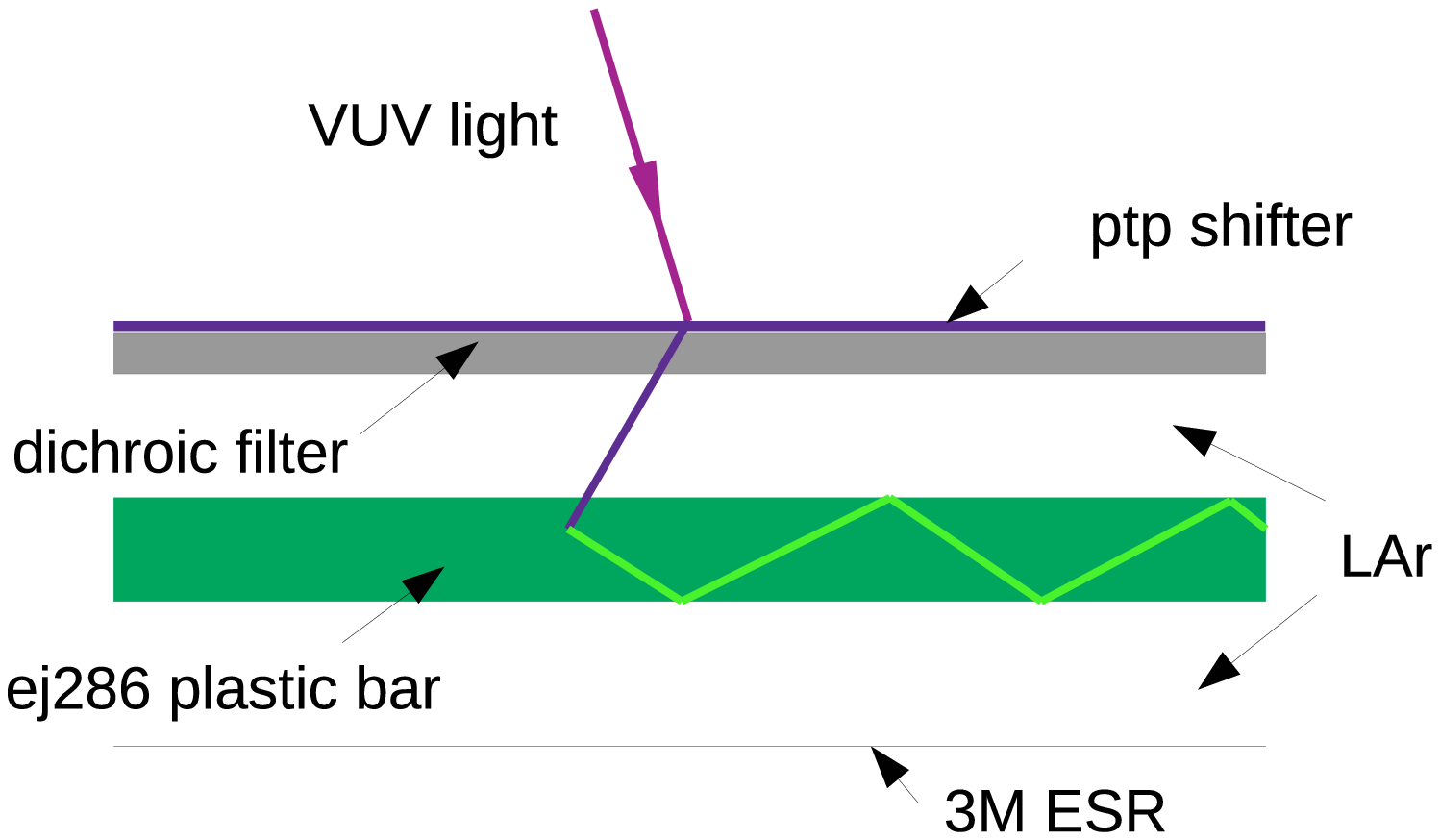}
        \includegraphics[width=0.49\textwidth]{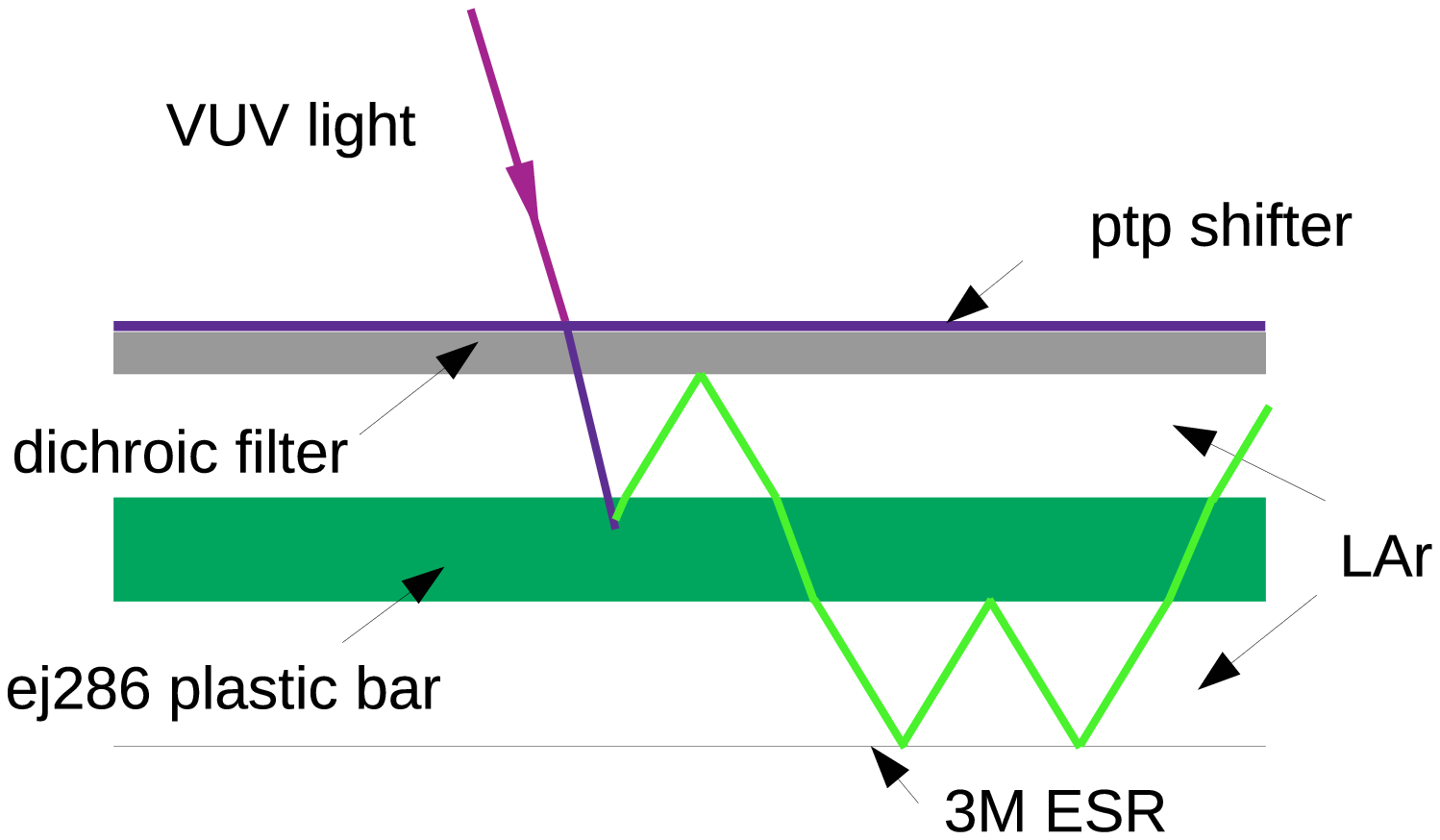}
    \caption{Schematic drawing of the X-ARAPUCA, with indication of its components and possible ways of photon trapping inside the cavity: through total internal reflection inside the WLS bar (left) or trapping by the ARAPUCA's cavity (right).}
    \label{fig:Esquemaxara}
\end{center}
\end{figure}

An improvement on the ARAPUCA device, the X-ARAPUCA \cite{xarapuca}, was proposed by replacing the second deposit of WLS by a WLS acrylic bar inside the box, and by having the SiPMs placed on the sides of the box. The photon entering the cavity could be trapped by the bar itself by total reflection or, if the angle is not shallow enough, by the original ARAPUCA trapping mechanism (see figure \ref{fig:Esquemaxara}).

The description of the ARAPUCA devices using Monte Carlo techniques has been used as a key tool as it allows design optimization, sensitivity studies, quantification of systematic errors, and development of analysis tools. It is therefore useful from the experimental proposal to the publication of results.

Building upon our previous experience in making estimates on the efficiency, number of reflections, and trapping time \cite{catania} of what is now called the standard ARAPUCA (S-ARAPUCA), we present a Monte Carlo modelling of the X-ARAPUCA device using the Geant4 framework \cite{geant, geant2}. The aim is to better understand the impact of different components on the detection efficiency of the device in order to guide the choices to be made for the Deep Underground Neutrino Experiment (DUNE) \cite{dune} since the X-ARAPUCA is the baseline for the photodetection system of the single-phase module. It is also going to be installed in the Short-Baseline Near Detector (SBND) \cite{sbnd}.

\section{Description of the Simulation}

The simulation is based on the Geant4 framework which provides the necessary functionalities for optical photon propagation and interaction with materials. 
For the description of the ARAPUCA device all of its components need to be carefully modelled considering their optical properties. The size of the box and number of SiPMs were varied to reproduce the devices to be installed in DUNE and SBND.

\begin{figure}[htbp]
    \centering
     \includegraphics[width=0.49\textwidth]{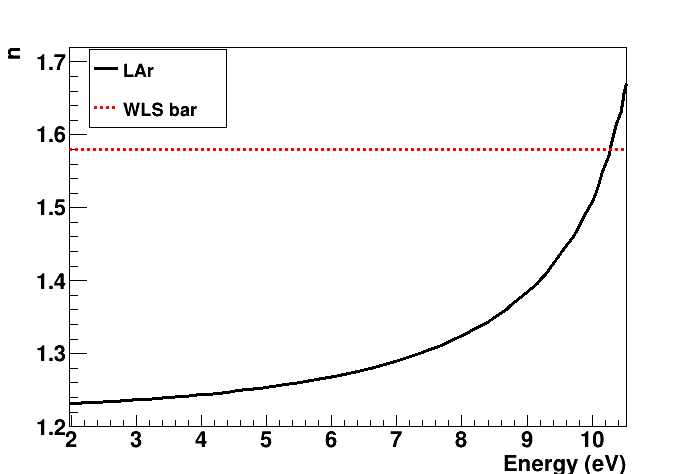}
     \includegraphics[width=0.49\textwidth]{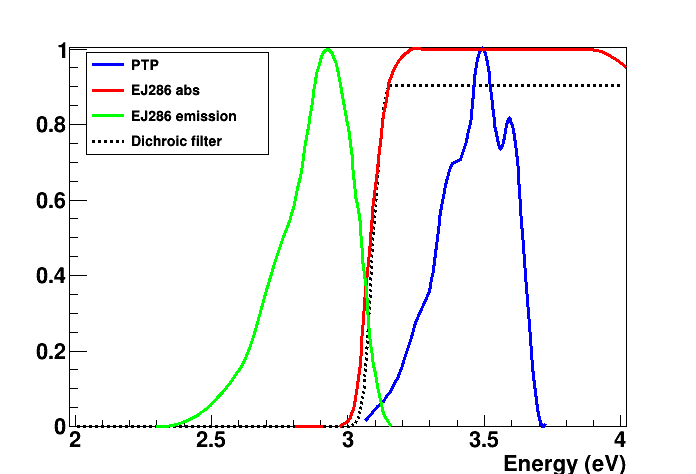}
     \includegraphics[width=0.49\textwidth]{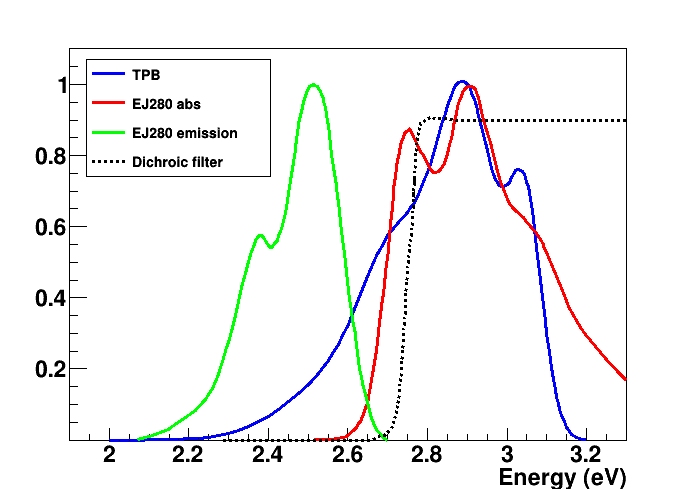}
      \includegraphics[width=0.49\textwidth]{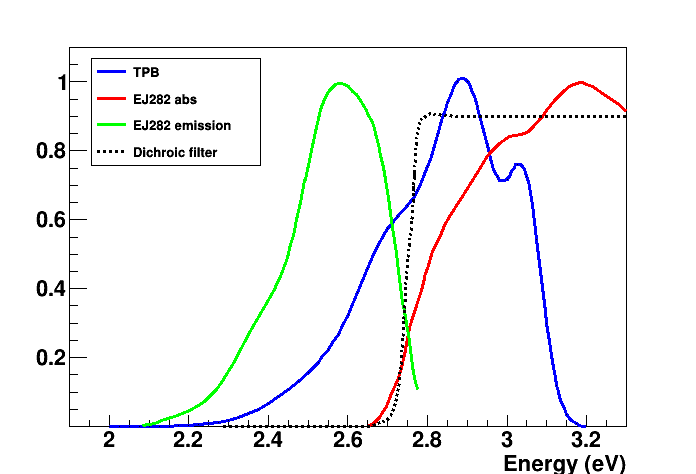}
    \caption{From top to bottom, left to right: (a) LAr refraction index as a function of photon energy and WLS bar refraction index adopted in the simulation; (b) Comparison between emission spectrum of PTP and absorption and emission of EJ286 (in arbitrary units), together with the dichroic transmission (at 45$^o$ angle and cutoff at 400 nm) as a function of the photon energy for detection of VUV light; Comparison between emission spectrum of TPB and absorption and emission of WLS bar (in arbitrary units), together with the dichroic transmission (at 45$^o$ angle and cutoff at 450 nm) as a function of the photon energy for detecting light converted by reflective foils when the bar is made of material (c) EJ280; and (d) EJ282.}
    \label{fig:coeff}
\end{figure}

The SBND experiment will have reflective foils composed of 3M ESR with TPB deposited over it installed on the cathode plane \cite{garcia2017}. The goal is to promote the conversion of LAr VUV light incident over the cathode allowing it to be reflected back to the photon detector region. The ARAPUCA, which is sensitive to the VUV LAr light, will not be able to detect the visible light already converted by the foils. In SBND there will two kinds of X-ARAPUCAS, one for detecting VUV and other for visible light. The former will have PTP deposited on the outer side of the dichroic filter with Eljen EJ286 as plastic bar for the internal wavelength shifter. The latter, will need no deposit over the filter, the photon being shifted only inside the box by the WLS bar. For this, two possibilities were considered: EJ280 and EJ282, whose
absorption and emission spectra cover the dichroic filter cutoff. The difference between their spectra can be seen in figure \ref{fig:coeff}. Simulation runs were performed with both in order to evaluate the best choice.

Table \ref{table:sim} shows the components used to model the X-ARAPUCA and main characteristics. Liquid argon present in the ARAPUCA internal cavity was modelled with a refraction index dependent on the photon energy as in reference \cite{lar}, reproduced in figure \ref{fig:coeff}. Details of the emission/absorption spectra of the shifters and comparison to the transmission coefficient of the dichroic filter for detecting VUV and visible light are also present in the same figure.

\begin{table}[htbp]
\begin{center}
\begin{tabular}{|l|p{11cm}|}
\hline
\textbf{Component} & \textbf{Characteristics} \\ \hline
3M ESR   & Taken as a specular reflector with reflectivity of 0.98 \cite{3m}. It covers the internal ARAPUCA's cavity surface.   \\ \hline
p-therfenyl   & Thickness of deposition of 2.0 um. Absorption length and emission spectrum dependent on photon energy from Ref. \cite{ptp}.  \\ \hline
WLS bar   &    Different bars from Eljen (EJ286, EJ280, and EJ282) to detect VUV or visible light. Absorption and emission spectra taken from manufacturer information as well as refraction index \cite{eljen}. 
\\ \hline
Dichoic filter   &    Transmission and reflection taken as a function of incident angle (information provided by manufacturer \cite{dicroico}). Dichroic filter placed over a glass substract and facing the ARAPUCA internal cavity. \\ \hline
SiPM & Detection spectrum taken as the one for Sensl C-Series \cite{sipm}. \\ \hline
\end{tabular}
\caption{ARAPUCA components and characterization in the simulation.}\label{table:sim}
\end{center}
\end{table}

\section{Results}

The devices were simulated in different configurations to evaluate the prototypes to be installed in SBND and the choices for DUNE. Given the uncertainties regarding the attenuation length in the WLS material, we compared the cases of no absorption of photons in the bar or an attenuation length of 1.0 meter.

\begin{figure}[htbp]
    \centering
     \includegraphics[width=0.49\textwidth]{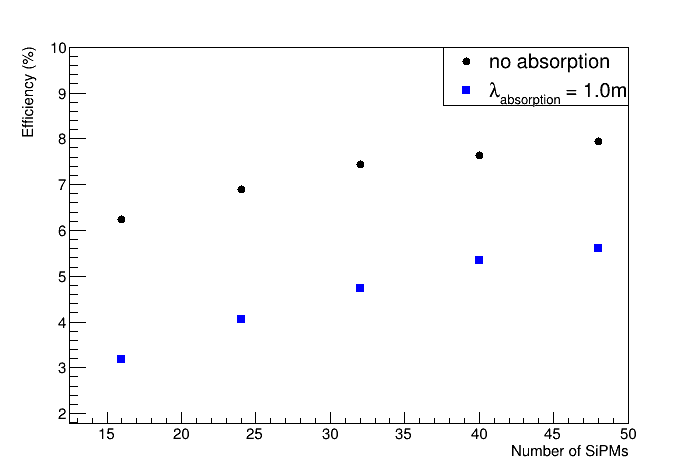}
     \includegraphics[width=0.49\textwidth]{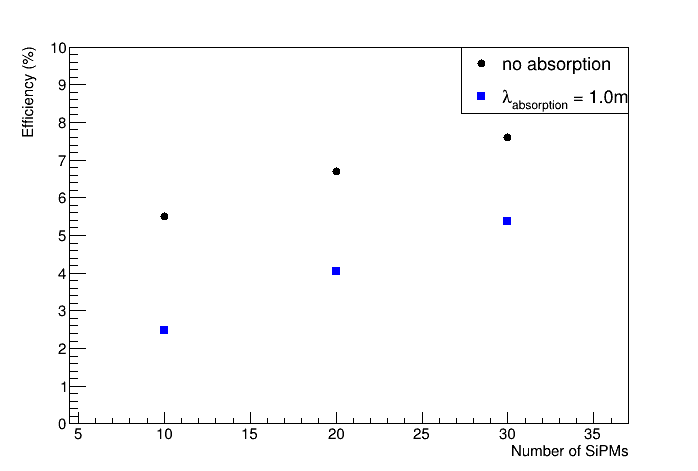}
     \includegraphics[width=0.49\textwidth]{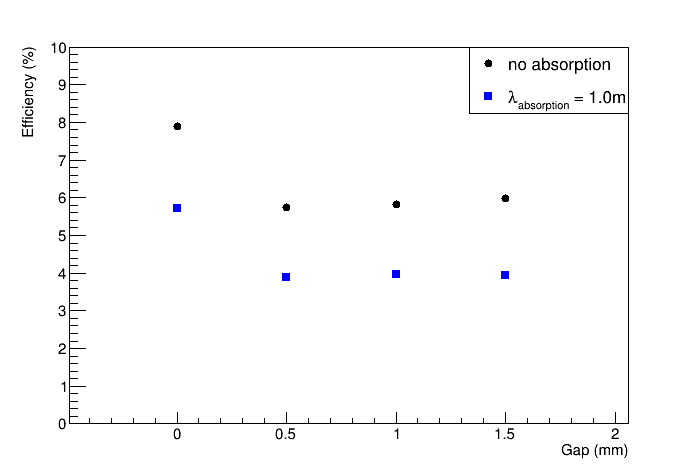}
      \includegraphics[width=0.49\textwidth]{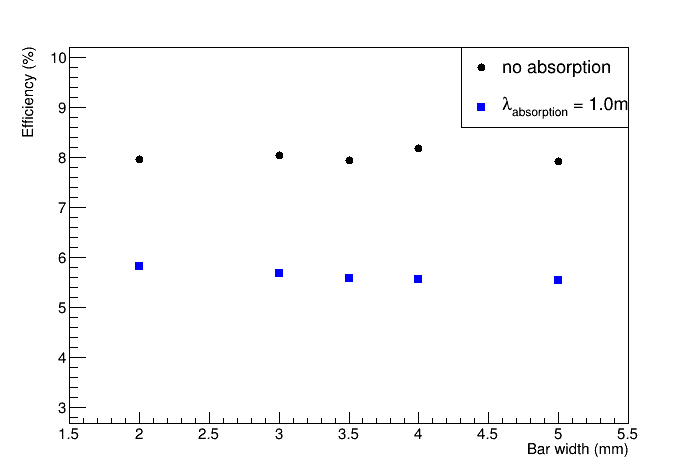}
    \caption{Efficiency studies for X-ARAPUCA as a function of (from top to bottom, left to right): (a) number of SiPMs equally spaced along the longer sides of the box; (b) number of SiPMs equally spaced on the shorter sides of the box; (c) spacing between the WLS bar and the SiPMs; (d) bar width. Results obtained for a device of internal size 480 mm x 93 mm x 6 mm, with 6x6 mm$^2$ SiPMs, and bar width of 3.5 mm, except when varied in case (d).}
    \label{fig:eff}
\end{figure}

\begin{figure}[htbp]
    \centering
    \includegraphics[width=0.99\textwidth]{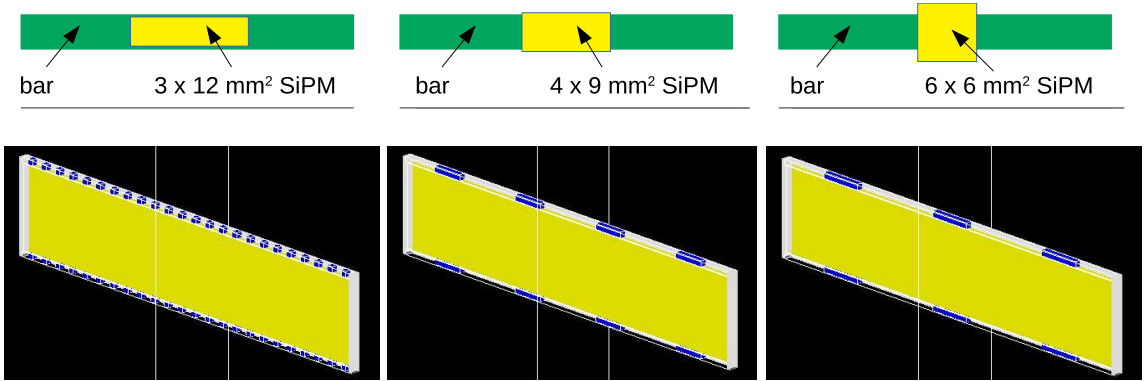}
    \caption{On top, simulating the effect of changing the geometry of SiPMs while keeping their surface areas. 
    Calculated efficiencies are, from left to right, 7.99$\pm$0.06\%, 8.12$\pm$0.06\%, and 7.95$\pm$0.06\%. On the bottom, simulating the effect of clustering SiPMs over the longer side of the X-ARAPUCA.
    Calculated efficiencies are, from left to right for SiPM clustered in groups of 1, 6, and 8, 7.95$\pm$0.06\%, 7.97$\pm$0.06\%, and 8.04$\pm$0.06\%, respectively.
    }
    \label{fig:sipm}
\end{figure}

Results can be seen in figures \ref{fig:eff} and \ref{fig:sipm} for a DUNE module (box internal dimensions of 480 mm x 93 mm x 6 mm). 

As expected, the efficiency for detection increases with the number of photon sensors but not linearly and with saturation
from 50 SiPMs. Simulation also indicates that the efficiency is improved by placing the SiPMs on the shorter side of the box.
This should be further investigated with experimental runs. 

One question investigated was related to the
shape of SiPMs, assuming the same coverage area, in order to improve
the detection efficiency. Figure \ref{fig:sipm} indicates that there is no need of having custom made SiPMs since the results are all compatible within the (statistical only) uncertainty. Also, the clustering of SiPMs seems to have no impact on the performance of the device, allowing for more degrees of freedom with regards to mechanical requirements of building and assembling of the box and readout electronics. We point out that these results were obtained without accounting for attenuation in the bar. 

Regarding properties of the bar, a thinner one would minimize the travel path but could also affect the conversion capability of the bar on PTP emitted photons. Nevertheless, simulation shows 
that the device is efficiently converting photons that enter the box and showing a slight efficiency reduction even when the absorption length is 1.0 m. The effect of absorption is to decrease the value of efficiency for all cases shown in figure \ref{fig:eff} in about 40-50\% for a 1.0 m value.

After assembly and installation of the devices in the experimental chamber, they will be immersed in liquid argon, at a temperature of $\sim$ 89K. Since component materials have different expansion coefficients, it is expected that a gap appears between the bar and the surface of the SiPM. Figure \ref{fig:eff}, third panel, shows the impact of this gap,
with a decrease in efficiency of about 25-30\%, independent on the gap
size. This is a very interesting result pointing to the possibility of moving the bar to be physically coupled to one of the box sides while only having a gap on the opposite side, thus the efficiency is reduced in a milder way. This might be accomplished by mechanically or gravitationally coupling one side of the bar to one lateral of the box.

Results for X-ARAPUCAS to be installed in SBND to detect visible light (box internal dimensions of 196.8 mm x 78.4 mm x 18 mm and 8 SiPM of 6 x 6 mm$^2$ of active area) are shown on table \ref{table:xarapuca}. As in the studies with the X-ARAPUCA prototype for DUNE, there is a decrease in efficiency with a decrease in the attenuation length and with the existance of a gap between the bar and SiPM surface. In the visible light X-ARAPUCA, the most interesting result from simulation was the comparison of the two bars, EJ280 and EJ282, where the former gives better results in efficiency due to the better match between the emission spectrum of TPB and absorption spectrum of EJ280 and emission of EJ280 and dichroic filter cutoff.

\begin{table}[htbp]
\begin{center}
\begin{tabular}{|l|c|c|c|}
\hline
\textbf{Bar type} & \textbf{Attenuation length} & \textbf{Space between bar and } & \textbf{Efficiency (\%)} \\ 
 & \textbf{(m)} & \textbf{lateral side (mm)} & \\ \hline
EJ280   &    $\infty$    & 1.5      & 3.5       \\ \hline
EJ280   & 1.0            & 1.5      & 2.2       \\ \hline
EJ282   &    $\infty$    & 1.5      & 2.6       \\ \hline
EJ282   & 1.0            & 1.5      & 1.9       \\ \hline
\end{tabular}
\caption{Results for X-ARAPUCA efficiency for light converted by TPB foils over SBND cathode. The box internal dimensions are 196.8 mm x 78.4 mm x 18 mm with 8 SiPMs of 6x6 mm$^2$ surface area uniformly distributed over the longer sides. Two types of WLS bars were tested with width of 4.0 mm.}\label{table:xarapuca}
\end{center}
\end{table}

\begin{figure}[htbp]
    \centering
     \includegraphics[width=0.49\textwidth]{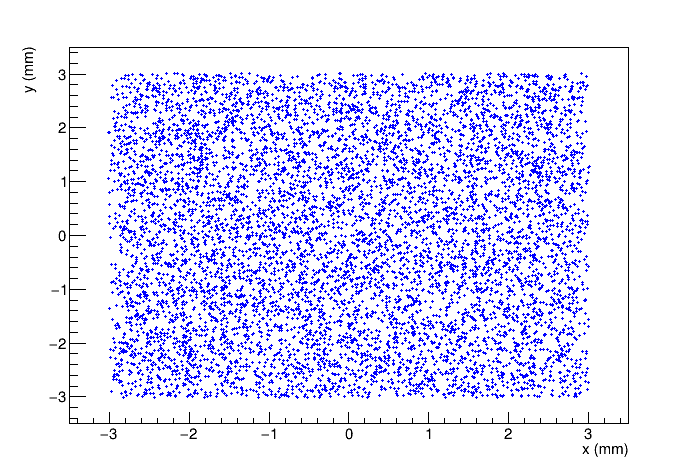}
     \includegraphics[width=0.49\textwidth]{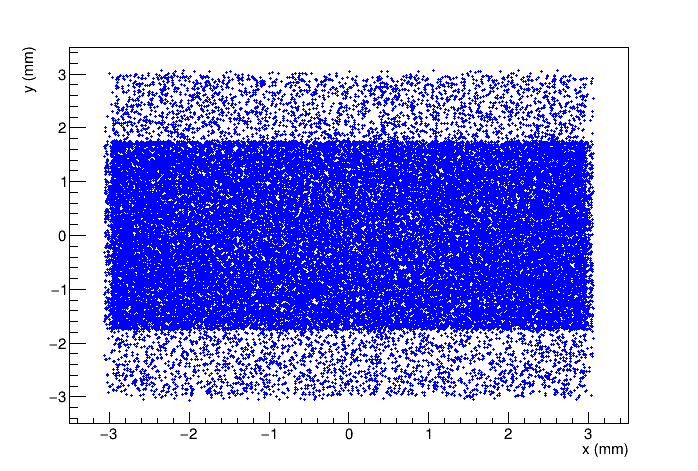}
    \caption{Comparison of the position over the SiPM area (6 mm x 6 mm) on which photons were incident for the S-ARAPUCA on the left and X-ARAPUCA on the right 
    (3.5 mm wide bar). The increase in the density of points over the area covered by the bar in the X-ARAPUCA is clear.}
    \label{fig:density}
\end{figure}

 When compared to a S-ARAPUCA of same size and number of SiPMs (the WLS in the bar is replaced by TPB deposited in the internal side of the dichroic filter), the simulation of a X-ARAPUCA device shows an efficiency improvement of 15-40\% (depending on properties of the bar and its coupling to the SiPMs). Figure \ref{fig:density} shows the positions over all SiPMs (of area 6x6 mm$^2$) where photons are incident on the sensor. It shows up to 55\% more photons in bar region, reflecting the fact that a large number of photons are being effectively trapped by the bar.

\section{Conclusions}

The development of the described simulation tools helps guiding design choices for the X-ARAPUCAs to be installed in the DUNE-SP \cite{dunesp} and SBND experiments. We highlight some interesting features such as the overall increase of the efficiency with respect to the previous device and its dependence with parameters such as the presence of a gap between the bar and SiPM regardless of its size and an improvement observed when placing the SiPMs on the shorter side of the cavity, a result that should be further investigated.

This simulation approach is flexible and can be adapted to different geometries,
resulting very helpful and effective in narrowing the choices for experimental setups where fewer configurations can be put to test guided by the simulation results.
It is also useful for investigating the impact of changes in single components over the device performance. ARAPUCA simulations have been validated against experimental data \cite{lnls, catania} proving to be a powerful tool in the device development.

\acknowledgments

The authors would like to thank Funda\c c\~ao de Amparo \`a Pesquisa do Estado de S\~ao Paulo (FAPESP) for financial support under grant no 2017/13942-5.

\end{document}